\documentclass[aip,reprint]{revtex4-1}
\usepackage{graphicx}
\usepackage{color}
\newcommand{\be}{\begin{equation}}
\newcommand{\ee}{\end{equation}}

\usepackage{bm}

\newcommand{\addcqt}{Centre for Quantum Technologies, National University of Singapore, 3 Science Drive 2, Singapore 117543}
\newcommand{\addtuc}{School of Electrical and Computer Engineering, Technical University of Crete, Chania, Greece 73100}

\begin{document}


\title{Dark soliton detection using persistent homology} 

\author{Daniel Leykam}
\email{daniel.leykam@gmail.com}
\affiliation{\addcqt}
\author{Irving Rond\'{o}n}
\affiliation{School of Computational Sciences, Korea Institute for Advanced Study, 85 Hoegi-ro, Seoul 02455, Republic of Korea}
\author{Dimitris G. Angelakis}
\affiliation{\addcqt}
\affiliation{\addtuc}

\date{\today}

\begin{abstract}
Classifying images often requires manual identification of qualitative features. Machine learning approaches including convolutional neural networks can achieve accuracy comparable to human classifiers, but require extensive data and computational resources to train. We show how a topological data analysis technique, persistent homology, can be used to rapidly and reliably identify qualitative features in experimental image data. The identified features can be used as inputs to simple supervised machine learning models such as logistic regression models, which are easier to train. As an example we consider the identification of dark solitons using a dataset of 6257 labelled atomic Bose-Einstein condensate density images.
\end{abstract}


\maketitle 

\begin{quotation}
The topological data analysis technique known as persistent homology characterizes complex high-dimensional datasets by studying how its ``shape,'' quantified by topological invariants, varies as a function of the scale at which the data is viewed. The ability of persistent homology to extract multi-scale information about the data, including its short range structure and long range global features makes it particularly promising for studying complex physical systems. So far, however, most applications of persistent homology to physics have focused on the analysis of theoretical models and numerically-generated data. Here we show how persistent homology can be applied to characterize experimental images of dark solitons in Bose-Einstein condensates, employing a large publicly-available density images. Persistent homology can rapidly and reliably detect topological features in the images related to the presence or absence of solitons. Combining persistent homology with simple machine learning models such as logistic regression can be used to automate the analysis and post-processing of large sets of experimental images.
\end{quotation}

\section{Introduction}

Machine learning techniques are attracting growing interest across the physical sciences as a means of automating analysis of complex experiments and discovering new features in high-dimensional datasets~\cite{ML_review,ML_photonics}. Many recent studies have employed deep artificial neural networks owing to their flexibility, carrying out diverse tasks including inverse design and image reconstruction~\cite{Feng2019,Pilozzi2018,Lim2020,Xu2020,Zhu2021,Haug2021,dataset}. However, neural networks are expensive to train (requiring extensive datasets and computing power) and it is challenging to interpret how exactly the trained models work and the conditions under which they may give wrong answers.

The limitations of neural networks motivate the use of more explainable machine learning techniques, including unsupervised learning and kernel methods~\cite{Ponte2017,Greitemann2019,Schuld2019,Che2020, Yeung2020}, which typically make use of some physical intuition behind the problem at hand. For example, kernel methods use nonlinear transformations to map the input data onto a feature space in which the data become linearly separable, such that simple clustering and classification algorithms work well~\cite{Schuld2019,Che2020}. This shifts the burden from training the model to finding suitable nonlinear feature maps. Commonly-used feature maps such as radial basis functions assume the data have some short range ordering or correlation length, but may fail to identify long range order or topological features such as loops and cavities.

Topological data analysis (TDA) is a powerful framework for identifying nonlocal features of high-dimensional datasets in a manner which is robust to noise and random perturbations~\cite{Ghrist2008,TDA_review,TDA_ML}. TDA was originally applied to problems in the life sciences, medicine, and data science. Interest in TDA methods is now growing among physicists~\cite{TDA_review2,Murugan2019}, with applications including the identification of amorphous phases of materials~\cite{Hiraoka2016}, classification of multiqubit entangled states~\cite{Mengoni2019}, time series analysis of chaotic systems~\cite{Robins2004,Maletic2016,Mittal2017,Tran2019,Tempelman2020,time_series}, detection of topological phase transitions~\cite{Donato2016,Tran2020, Olsthoorn2020,Cole_arxiv}, and more~\cite{Spitz2020,Rocks2021,Leykam2021,Bhaskar2019, Maletic2021}. However, applications of TDA to physics have largely focused on proofs-of-concept using synthetic datasets. We expect the robustness of TDA to really shine when applied to noisy experimental data. The objective of this study is to test the performance of TDA on a large experimental dataset.

Specifically, we consider a recently-released collection of 6257 atomic density images of Bose-Einstein condensates (BECs), labelled as containing no solitons, a single dark soliton, or multiple excitations~\cite{dataset}. The authors of Ref.~\cite{dataset} constructed this dataset to train a convolutional neural network to distinguish the three classes of images. The neural network classifier exhibited performance comparable to a human classifier, enabling quick automatic processing of experimental images using a standard laptop computer. The dataset of Ref.~\cite{dataset} provides an ideal testbed for the TDA-based analysis of data arising from experiments studying solitons and other nonlinear waves.

In this article we test a TDA-based approach for high-throughput classification of experimental image data, considering the task of dark soliton identification. Specifically, we study the persistent homology of the low density regions of atomic BEC images. Persistent homology computes the range of intensity scales over which topological features of images, such as their local intensity minima and low intensity lines, exist. Dark solitons can thereby be identified in one-dimensional (1D) and two-dimensional (2D) images as highly persistent minima and low intensity lines, respectively. We show that simple point summaries of these features' persistence -- their entropy and $p$-norms -- can be used to distinguish soliton-containing from soliton-free images with reasonable accuracy in a manner robust to noise-induced local density minima.

The outline of this article is as follows: Sec.~\ref{sec:method} provides a brief introduction to the dataset and TDA-based image analysis using persistent homology. Sec.~\ref{sec:1D} presents an approach for rapidly identifying dark solitons in 1D density images using their persistent zero-dimensional features (distinct low intensity regions). Sec.~\ref{sec:2D} generalizes the method to the full 2D density images using persistent 1D features (lines of low intensity), which has the advantage of not requiring the image orientation to be known a-priori. Sec.~\ref{sec:discussion} discusses the performance of our approach compared to the convolutional neural network of Ref.~\cite{dataset}. Sec.~\ref{sec:conclusion} concludes.

\section{Approach}
\label{sec:method}

The dataset of Ref.~\cite{dataset} consists of 6257 condensate density images, labelled as containing no solitons (1237 images), a single dark soliton (3468 images), or other excitations (1552 images). Each image has been cropped and centred by fitting the pixel values to a 2D Thomas-Fermi distribution,
\be 
n^{\mathrm{TF}}(x,y) = n_0 \mathrm{max}\left[ 1 - (\frac{x-x_0}{r_x})^2 + (\frac{y-y_0}{r_y})^2, 0 \right]^{3/2} + \delta n,
\ee 
where $n_0$ is the peak density, $\delta n$ is an offset, the centre of the BEC lies at coordinates $(x_0,y_0)$, and $r_{x,y}$ are the Thomas-Fermi radii. $\delta n = 0.25$ in the provided images. As a pre-processing step, we apply a uniform shift to the pixel values such that $\delta n \rightarrow 0$, and then rescale the images such that each has a peak density (intensity) normalized to 1. 

Assuming any dark solitons are much smaller than the total condensate size and are propagating along the $x$ axis, their density profile at time $t$ is locally approximated by the profile of a 1D dark soliton on a uniform background $n$. Using dimensionless units in which density, time, and length are normalized by the $s$-wave scattering length, trapping period, and trapping length, respectively, the dark soliton profile takes the form~\cite{dark_soliton_review,soliton_book}
\be 
n_{\mathrm{DS}}(x) = v^2 + [n - v^2] \tanh^2 [ w(x - vt) ], \label{eq:soliton}
\ee 
where $v$ is the normalized soliton velocity and $w = \sqrt{n - v^2}$ is its width. The density minimum of $n_{\mathrm{min}} = v^2$ occurs at $x=vt$. The soliton has power (integrated density depletion) $P_r = 2 w = 2 \sqrt{n - n_{\mathrm{min}}}$ and energy $H_r = \frac{1}{6} P_r^3$. Thus, the density contrast in the vicinity of the soliton core provides a measure of its power, or alternatively the significance of the density depletion. Note that Eq.~(\ref{eq:soliton}) is the soliton profile in an infinite 1D system; deviations will occur when there is an additional confining potential, as is the case here.

The goal is to distinguish significant dark soliton-induced density depletions from local minima caused by density fluctuations of the condensate and imaging noise. If the number of solitons and their orientation is known in advance, a simple and effective way to identify their positions is by performing least squares fitting of the dips to Gaussian functions~\cite{Fritsch2020}. We desire a method which does not require manual identification of the number of solitons or knowledge of their orientation.

\begin{figure}
    \centering
    \includegraphics[width=\columnwidth]{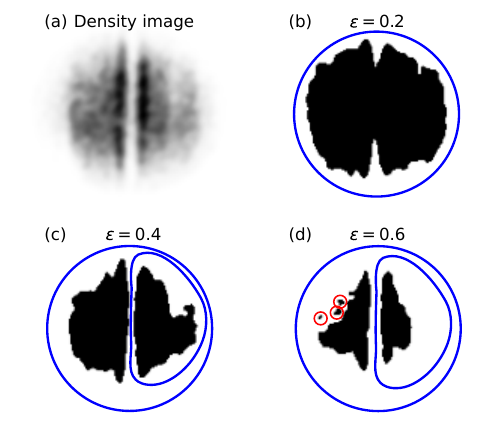}
    \caption{(a) Sample image after normalization and smoothing using a Gaussian filter with standard deviation $\sigma = 2$. (b,c,d) Low intensity regions sets for different cutoff intensities $\epsilon$. Black indicates pixels above the threshold. (b) For small $\epsilon$ only the exterior of the condensate is included, corresponding to a low intensity region containing a single cycle, indicated in blue. (c) The soliton density depletion results in an additional long-lived cycle once $\epsilon$ exceeds a critical value. (d) Small-scale density fluctuations can generate additional short-lived cycles that persist only over a small range of cutoff intensities, indicated in red.}
    \label{fig:filtration_example}
\end{figure}

Our approach is to use a TDA method known as persistent homology, which studies the persistence of topological features of datasets over a range of characteristic scales~\cite{Ghrist2008,TDA_review,TDA_ML}. 

In the case of grayscale image data, the first step is to convert it to a binary image using an intensity threshold $\epsilon$; all pixels with intensity values less than $\epsilon$ are rounded to zero, while those with intensites above $\epsilon$ are rounded to one. Then, we consider the topology of the zeroed pixels, i.e. the number of disconnected components (clusters) and non-contractible lines (loops). Fig.~\ref{fig:filtration_example} illustrates this procedure for a sample image containing a single dark soliton. As the threshold $\epsilon$ is increased, low density ``valleys'' of the image are progressively filled. Topological features (clusters, cycles) are created or destroyed at critical values $\epsilon$ corresponding to local minima, maxima, or saddle points of the intensity. This serves as a reliable way of segmenting images, particularly when the shapes of the individual segments may be highly irregular.

Persistent homology characterizes images by tracking the intensity scales $\epsilon$ at which topological features appear (birth scale $b$) and are destroyed (death scale $d$) to create a persistence diagram $D$, which forms a topological fingerprint of the image. Importantly, persistence diagrams for low-dimensional topological features such as clusters and loops can be computed efficiently~\cite{Ripser,Ripser_py,scikit-tda}. Tracking the persistence of each feature allows to distinguish significant dark soliton-induced minima from noise-induced minima with low persistence. Once a persistence diagram has been computed from an image there are a variety of methods for passing the encoded information into standard machine learning techniques. 

Motivated by the fact that the soliton power depends only on the density contrast, and not the absolute values of the background density and density minima, we will use the feature persistence $|d-b|$ (also known as feature lifetime) to characterize the images. Specifically, we will use the number of features with anomalously large persistence to estimate the number of dark solitons in the images. We employ two summary statistics to characterize the distribution of feature persistences $|d-b|$ and detect the presence of features with anomalously large persistence. 

The first summary statistic is the persistent entropy~\cite{entropy1,entropy2,entropy3,entropy4},
\be 
\mathcal{E}(D) = -\sum_{(b,d)\in D} \frac{|d-b|}{\mathcal{S}(D)} \log \left( \frac{|d-b|}{\mathcal{S}(D)} \right), \label{eq:entropy}
\ee 
where $\mathcal{S}(D) = \sum_{(b,d)\in D} |d-b|$ is the sum of feature persistences in the persistence diagram $D$ (excluding any features with infinite persistence); the summation is over all points (features) of the dimension of interest appearing in a persistence diagram. The persistent entropy measures the disorder in the distribution of feature persistences; noise-induced features tend to have similar persistence, corresponding to larger values of $\mathcal{E}$. $\mathcal{E} = 0$ if there is only a single feature, and $\mathcal{E} = \mathrm{log}(N)$ if the diagram consists of $N$ features, all with the same persistence.

The second summary statistic we consider is the $p$-norm of the feature persistences~\cite{Tran2020,stability}
\be 
\mathcal{P}_p(D) = \left(\sum_{(b,d)\in D} |d-b|^p \right)^{1/p}. \label{eq:pnorm}
\ee
For example, $\mathcal{P}_{\infty}$ measures the persistence of the longest-lived feature in $D$, $\mathcal{P}_1$ is the sum of feature persistences, $\mathcal{P}_2$ is the root mean squared persistence, and so on.

Computing the summary statistics Eqs.~(\ref{eq:entropy}) and~(\ref{eq:pnorm}) reduces the complex, high-dimensional image data into a more easily interpretable set of low-dimensional feature vectors; each image is mapped to point in this low-dimensional space. The standardised feature vectors (i.e. with components rescaled to have zero mean and unit variance) can then be used as inputs to the machine learning model of choice. In the following will consider a simple linear classifier model (logistic regression), which assigns each image a probability of belonging to a particular class. We use 5\% of the images to train the logistic regression model and test the its performance at classifying the remaining images~\cite{scikit-learn}.

We use the $F_1$ score to quantify the model performance, and to enable comparison with the results of Ref.~\cite{dataset}. The $F_1$ score $\in [0,1]$ is the harmonic mean of the classifier's precision and recall, which relate to the number of false positives and false negatives returned, respectively. A higher $F_1$ score indicates better performance at classifying the test images.

\section{1D image analysis}
\label{sec:1D}

\begin{figure}
    \centering
    \includegraphics[width=\columnwidth]{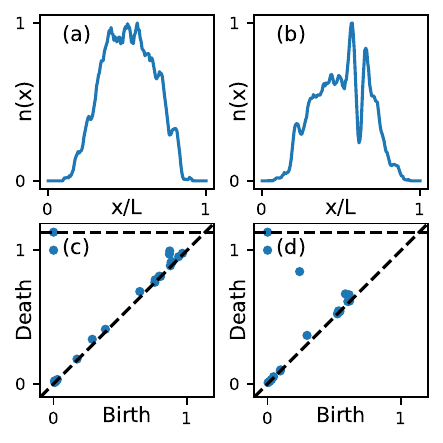}
    \caption{(a,b) Sample 1D density profiles from the BEC image dataset, labelled as containing no solitons (a) and a single soliton (b), the latter visible as a prominent density dip. (c,d) Persistence diagrams illustrating intensity scales at which low intensity clusters are created (birth) and destroyed (death).}
    \label{fig:1D_filtrations}
\end{figure}

\begin{figure}
    \centering
    \includegraphics[width=\columnwidth]{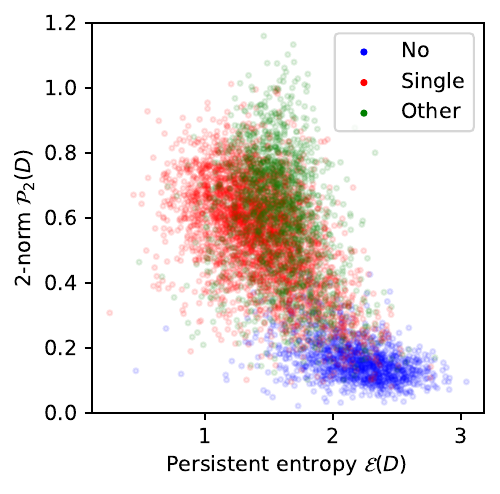}
    \caption{Separation of the three image classes using the two-dimensional feature space formed by the persistent entropy $\mathcal{E}$ and $2$-norm of the feature persistences $\mathcal{P}_2$. Each point corresponds to a single image of the full dataset, with the colour indicating its label.}
    \label{fig:1D_features}
\end{figure}

As an introductory example, we consider the simpler case of 1D density profiles obtained by integrating the images over the $y$ axis, $n_{\mathrm{1D}}(x) =  \sum_{y} n(x,y)$. We rescale the density such that $\mathrm{max}(n_{\mathrm{1D}}) = 1$ in all images. Fig.~\ref{fig:1D_filtrations}(a,b) shows density profiles for two sample images labelled as containing no solitons and a single soliton, respectively. By computing the connectivity of the low intensity images as a function of the cutoff intensity $\epsilon$ we obtain the persistence diagrams shown in Fig.~\ref{fig:1D_features}(c,d), which plot the birth and death intensity scales of the low density clusters. Note that for 1D images the only topological features are clusters; there are no loops.

Both persistence diagrams have many points close to the diagonal, corresponding to short-lived noise-induced low intensity clusters which are destroyed shortly after they appear. In the upper left corners of the persistence diagrams there is a pair of long-lived features, corresponding to the density minima at the edges of the condensate. Note that the dashed horizontal line is used to indicate features with an infinite persistence. The single infinite persistence feature reflects the fact that once the threshold exceeds the maximum pixel value, the entire image forms a single low intensity cluster. The second image, labelled as containing a single dark soliton, exhibits an additional high persistence minimum corresponding to the soliton-induced density depletion.

Before computing the summary statistics of the persistence diagrams of the 1D density profiles, we exclude their two longest-lived features from $D$, since these originate from the exterior of the BEC and are common to all images in the dataset. Then, images with no persistent minima (i.e. no dark solitons) will only have noise-induced features lying close to the diagonal, corresponding to a high entropy $\mathcal{E}$ and low $\mathcal{P}_p$. On the other hand, images containing solitons will have long-lived features resulting in smaller entropy and larger $\mathcal{P}_p$. 

We compute $\mathcal{E}$ and $\mathcal{P}_2$ for all images in the dataset and plot them in Fig.~\ref{fig:1D_features}, coloured according to the image label. Computation of low intensity clusters of 1D images can be reduced to sorting the pixels by their value, and can thus be performed quickly; computation of the summary statistics required about 1 ms/image using a standard laptop computer. The two summary statistics are sufficient to linearly separate images containing no solitons from the other images, while it is harder to distinguish single solitons from images containing other excitations.

\begin{figure}
    \centering
    \includegraphics[width=\columnwidth]{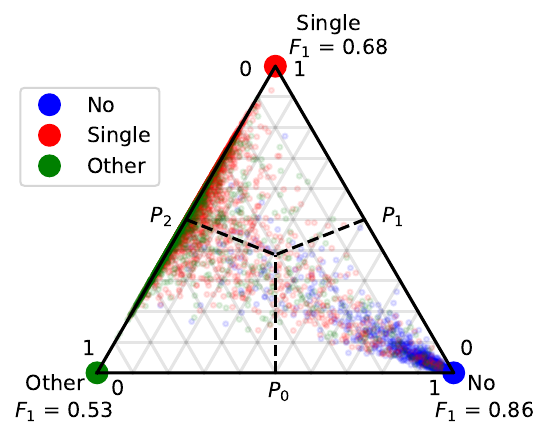}
    \caption{Performance of a logistic regression classifier at distinguishing the three image classes using the persistent entropy $\mathcal{E}$ and the feature persistence $2$-norm $\mathcal{P}_2$. Dashed lines indicate decision boundaries. $F_1$ scores $\in [0,1]$ for each class are shown at the triangle corners (the higher the better).}
    \label{fig:1D_test}
\end{figure}

Fig.~\ref{fig:1D_test} shows the results of the logistic regression classifier on the full test set. As expected from Fig.~\ref{fig:1D_features}, the soliton-free images are more confidently distinguished from the images with solitons, while the classifier has lower confidence at distinguishing single solitons from images with multiple excitations, with most of the latter images lying close to the decision boundary. The lower $F_1$ score for the ``Other excitations'' class in Fig.~\ref{fig:1D_test} indicates the classifier struggles to correctly identify this class.

\section{2D image analysis}
\label{sec:2D}

The use of 1D density profiles in the previous Section assumes that the dark stripe associated with the solitons is parallel to the vertical axis. This is similar to the convolutional neural network approach of Ref.~\cite{dataset}, which requires the images to be a aligned as a pre-processing step. To overcome this limitation, we now consider the full 2D density images. 

\begin{figure}
    \centering
    \includegraphics[width=\columnwidth]{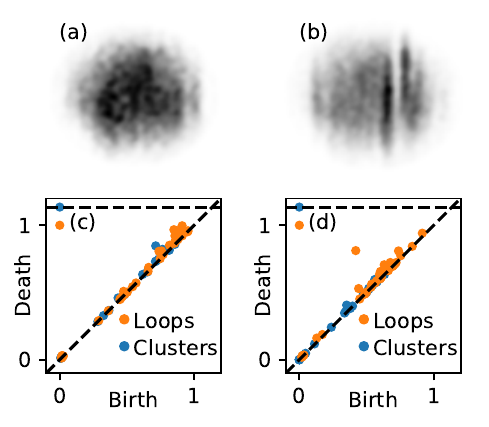}
    \caption{(a,b) 2D BEC density images labelled as containing no solitons (a) and a single soliton (b) after performing spatial smoothing to filter out rapid density fluctuations. (c,d) Corresponding persistence diagrams illustrating intensity scales at which different features corresponding to distinct low intensity clusters (blue) and lines (brown) are created (birth) and destroyed (death).}
    \label{fig:2D_example}
\end{figure}

In Fig.~\ref{fig:2D_example} we compute the persistence diagrams of the images from Fig.~\ref{fig:1D_filtrations}, this time without summing over $y$. As a preprocessing step we smoothed the images by applying a Gaussian filter with standard deviation $\sigma = 2$ pixels to reduce the number of features with very short persistence.

In the 2D images there is always only a single cluster with a significant persistence, reflecting the fact that any low-density regions associated with dark solitons should cross the entire cloud and be connected to the low density exterior. Therefore, we focus on the persistence of 1D features (low intensity loops) to identify dark solitons. In this case, there is always a single long-lived loop (corresponding to the BEC cloud acting as a large ``hole'' in the low density exterior region), which we exclude from the following analysis. Then for every dark soliton in the image there will be an additional long-lived loop appearing in the persistence diagram.

We computed the two summary statistics $\mathcal{E}$ and $\mathcal{P}_2$ of the loop feature persistences, similar to the previous Section. While the two summary statistics can clearly distinguish images containing solitons from the soliton-free images, the single dark soliton and other excitations classes are not clearly separated. This leads to a classifier performance significantly worse than the classifier based on the 1D density images. The poorer performance is likely due to the large number of lower persistence, noise-induced features seen in Fig.~\ref{fig:2D_example}, which lowers the sensitivity of the measures $\mathcal{E}$ and $\mathcal{P}_2$ to the few soliton-induced features. 

\begin{figure}
    \centering
    \includegraphics[width=\columnwidth]{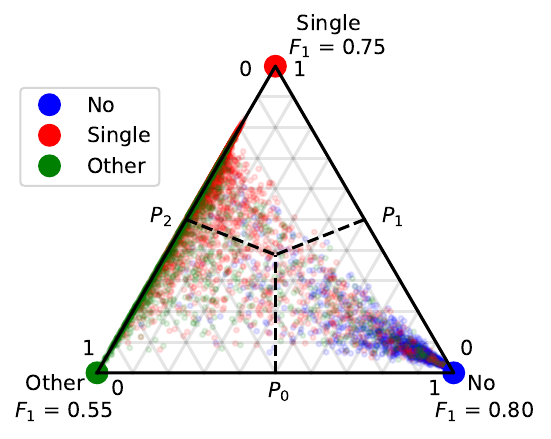}
    \caption{Performance of a logistic regression classifier at distinguishing the three image classes using the six-dimensional feature vector $(\mathcal{E},\mathcal{P}_1, \mathcal{P}_2, \mathcal{P}_3,\mathcal{P}_{4}, \mathcal{P}_{\infty})$ obtained from the loop features of the 2D density images.}
    \label{fig:2D_test}
\end{figure}

The influence of the noise-induced features can be reduced by either excluding features with persistence below a chosen threshold, or introducing summary statistics summaries sensitive to outliers in the persistence distribution. We consider the latter, computing additional $p$-norms to construct a 6-dimensional feature vector $(\mathcal{E},\mathcal{P}_1, \mathcal{P}_2, \mathcal{P}_3,\mathcal{P}_{4}, \mathcal{P}_{\infty})$. The performance of this improved logistic regression model is illustrated in Fig.~\ref{fig:2D_test}. In particular, the individual $F_1$ scores becomes closer to those obtained using the 1D image analysis method. 

\section{Discussion}
\label{sec:discussion}

We summarize the performance of our trained logistic regression classifiers in Tab.~\ref{tab:scores}, including binary classifier models trained to only distinguish the soliton-free images from the two other image classes. The additional $p$-norm summary statistics mainly serve to improve the accuracy at distinguishing the single soliton and other excitation classes in the 2D images, yielding only a marginal improvement for the 1D images and the soliton-free class.

\begin{table*}
\centering
\begin{tabular}{|c|c|c|c|c|}
\hline
     & \multicolumn{2}{|c|}{1D images}  & \multicolumn{2}{|c|}{2D images} \\
\hline
Features used & ($\mathcal{E},\mathcal{P}_2$) & ($\mathcal{E},\mathcal{P}_{1-4,\infty}$) & ($\mathcal{E},\mathcal{P}_2$) & ($\mathcal{E},\mathcal{P}_{1-4,\infty}$) \\
\hline
No soliton F1 & 0.86 & 0.87 & 0.79 & 0.80 \\
Single soliton F1 & 0.68 & 0.76 & 0.60 & 0.75 \\
Other excitations F1 & 0.53 & 0.61 & 0.45 & 0.55 \\
Binary weighted F1 & 0.94 & 0.94 & 0.91 & 0.91 \\
\hline
\end{tabular}
\caption{Summary of the performance of the logistic regression models considered, quantified by $F_1$ scores $\in [0,1]$ (the higher the better). ``Binary weighted'' indicates the performance at distinguishing the ``No soliton'' images from the ``Single soliton'' and ``Other excitations'' classes.}
\label{tab:scores}

\end{table*}

For comparison, the best neural network of Ref.~\cite{dataset} achieved $F_1$ scores of 0.96, 0.91, and 0.81 for the no soliton, single soliton, and other excitation classes, respectively. However, this higher performance required time-consuming optimization of the neural network hyperparameters, with many of the neural network models performing significantly worse.

One big advantage of the TDA approach is that the topological features employed are invariant under image rotations. We repeated the 2D image analysis and model training with each image rotated by a random angle before processing, observing identical performance. In contrast, image analysis using convolutional neural networks typically requires either alignment as a preprocessing step, or augmenting the training set with rotated images. 

The neural network models of Ref.~\cite{dataset} also had difficulty with reliably identifying the ``Other excitations'' class, achieving a moderate $F_1$ score of 0.81. There are two reasons for this low score. Firstly, some images in this class are genuinely difficult to classify, even for expert humans. Secondly, this class includes images with various soliton numbers occurring too rarely for accurate supervised classification. One direction for future research will be to use interpretable features such as $p$-norms to perform unsupervised clustering of the images to distinguish different numbers of dark solitons. 

It interesting to note that already a few point summaries are sufficient for moderate accuracy classification of the BEC image; our analysis discarded a lot of the information encoded in the persistence diagrams. There are a variety of stable distance measures and vectorizations for comparing persistence diagrams without loss of information, including the Wasserstein distance and persistence landscapes~\cite{TDA_review,TDA_ML,TDA_review2}. We initially tried to train our linear classifier using these distance measures, but did not obtain good performance. One possible reason is that these measures are sensitive to the feature birth and death times, whereas for soliton identification we are mainly interested in the feature persistence. One may be able to improve the performance by feeding the features obtained using persistent homology into a nonlinear model, such as an artificial neural network.

\section{Conclusion}
\label{sec:conclusion}

In summary, we have proposed persistent homology-based approaches for high-throughput identification of dark solitons in a dataset of Bose-Einstein condensate density images. Persistent homology characterizes the images by computing the topological properties (number of clusters and loops) of the low intensity regions of the images as a cutoff between low and high intensity regions is varied. The range of intensity scales over which different clusters and loops persist provides a topological fingerprint of each image in the form of a persistence diagram.

We showed that a few simple summary statistics of the persistence diagrams -- their entropy and $p$-norms -- can serve as features that linearly separate images with solitons from soliton-free images, while being insensitive to noise-induced local density minima. The first approach we considered is tailored to 1D images and can distinguish soliton-free images from images containing solitons with high confidence. The second approach, based on persistence diagrams of 1D features has a comparable accuracy, but does not require the image orientation to be known in advance. 

Our approach can be applied to a wide variety of object detection problems, including the detection of bright solitons, which may be identified as persistent local density maxima. We hope that our findings illustrate how suitably-designed interpretable machine learning approaches can offer a powerful alternative to deep neural networks, particularly in applications where computing power or time are limited. 

\begin{acknowledgments}
This research was supported by the National Research Foundation, Prime Minister's Office, Singapore, the Ministry of Education, Singapore under the Research Centres of Excellence programme, the Polisimulator project co-financed by Greece and the EU Regional Development Fund, and the Basic Science Research Program through the National Research Foundation of Korea (NRF), funded by the Ministry of Science and ICT (NRF- 2017R1E1A1A01077717).
\end{acknowledgments}

\section*{Author Declarations}

\subsection*{Conflict of Interest}

The authors have no conflicts to disclose.

\subsection*{Data availability}

The code used to perform the numerical calculations within this paper is available from the corresponding author upon reasonable request.


\end{document}